\documentclass[twocolumn,pra]{revtex4}
\usepackage{graphicx}

\begin{document}

\title{Entanglement and spin squeezing of Bose condensed atoms}

\author{M. Zhang$^1$, Kristian Helmerson$^2$, and L. You$^1$}
\affiliation{$^1$School of Physics, Georgia Institute of
Technology, Atlanta, GA 30332-0430}
\affiliation{$^2$Atomic Physics Division,
National Institute of Standards and Technology,
Gaithersburg, MD 20899-8424}

\date{\today}

\begin{abstract}
We propose a direct, coherent coupling scheme that can create
massively entangled states of Bose-Einstein condensed atoms. Our
idea is based on an effective interaction between two atoms from
coherent Raman processes through a (two atom) molecular
intermediate state. We compare our scheme with other recent
proposals for generation of massive entanglement of Bose condensed
atoms. We also present explicit calculations that support
detecting maximally entangled states without requiring single
atom resolved measurements.
\end{abstract}

\pacs{03.65.Ud 03.65.Bz 42.50.-p 03.75.Fi}

\maketitle

\section{introduction}
Entanglement lies at the heart of the difference between the
quantum and classical multi-particle world. It is the phenomenon
that enables quantum information and quantum computing with many
qubits. Recently, several interesting developments have occurred
in studies of massively entangled atomic states. Based on the
proposals of Sorensen and Molmer \cite{sorensen99}, a controlled,
entangled state of $4-$ions was successfully created by the NIST
ion trap group \cite{sackett00}. Zeilinger and coworkers prepared
three entangled photon or GHZ states by selecting from two beams
of entangled photon pairs \cite{bouwmeester99}. Entanglement
between two atoms and a microwave photon were also detected in a
"step-by-step" process \cite{rauschenbeutel00}.

Of these and other related developments, the idea of Molmer and
Sorensen \cite{sorensen99} is especially interesting. In their
scheme a direct coupling to the multiparticle, entangled final
state was made possible through a virtual, intermediate state
which was a common (quantum) mode of the motion of all the ions.
Similar type interactions were also obtained by Milburn
\cite{milburn99}. Both proposals allow for creation of massive
entangled states by unitary evolution, starting from certain pure
initial states.

A Bose-Einstein condensate is a convenient source of atoms in
initially pure and separable states.  Zoller and coworkers
suggested creating massive entangled spin squeezed states from a
two component condensate using the inherent atom-atom interactions
\cite{sorensen01} and further investigated schemes to dynamically
create many particle entangled states of a two component BEC
\cite{micheli02}. Spin exchange interaction of a spinor
condensate was also proposed as a candidate for creating entangled
pairs of atoms \cite{pu0007,duan0011}.

Most of these proposals work in the two mode approximation where
one motional state is assumed for each spinor component of
condensed atoms. Recently, Sorensen demonstrated the validity of
this method within Bogoliubov theory \cite{sorensen02}. Similar
studies were previously performed, extensively, for condensate
atoms in a double well potential
\cite{Javanainen99,zapata98,spekkens99,imamoglu97,milburn97,ho00}.

Recently, two of us proposed a new type of atom-atom coupling that
achieves optimal spin squeezing in two mode Bose condensed atoms
\cite{helmerson01}. In this study we provide further details of
our proposal and compare its potential for both the creation of
massive entangled states \cite{sorensen99} and improved spin
squeezing \cite{sorensen01,pu0007,duan0011,kitagawa93}. This paper
is organized as follows. First, we briefly review the Raman
process with photon momentum transfer for a single three-level
$\Lambda$-type atom. We then extend the analysis to the case of
two atoms, using intermediate molecular states. The resulting
effective interaction is then compared with models in several
other recent studies of creating entanglement with Bose condensed
atoms. We then provide an explicit calculation of the time
reversed dynamics to confirm the generation of maximally
entangled states without requiring single atom resolved
measurements \cite{micheli02,vogels02}.
We conclude with a careful analysis of the limitations and
advantages of our model.

\section{Single atom, two-photon process
through an atomic intermediate state}
We consider a three level $\Lambda$-type atom described by the
Hamiltonian
\begin{eqnarray}
H=\sum_{\mu=g,e,g'}\left[\frac{p^2}{2M}+V_{\mu}(\vec{r})+\hbar\omega_{\mu
g}\right]\left|\mu\right\rangle\!\left\langle\mu\right|+H_{{\rm int}},
\end{eqnarray}
where the dipole interaction term is
\begin{eqnarray}
H_{{\rm int}}&=&-\vec{d}_{eg}\cdot\vec{E}_1(\vec{r})
e^{i\vec{k}_1\cdot\vec{r}-i\omega_1t}
\left|g\right\rangle\!\left\langle e\right|+h.c.\nonumber\\
&&-\vec{d}_{eg'}\cdot\vec{E}_2(\vec{r})
e^{i\vec{k}_2\cdot\vec{r}-i\omega_2t}\left|g'\right\rangle\!\left\langle
e\right|+h.c.
\end{eqnarray}
for the general Raman process from initial state
$\left|g\right\rangle$ to final state $\left|g'\right\rangle$
through an intermediate state $|e\rangle$.
$V_{\mu}(\vec{r})$ denotes the, possibly different, trapping
potentials for the different electronic states
$\left|\mu\right\rangle$ with internal energy $\omega_{\mu
g}=\omega_{\mu\mu}-\omega_{gg}$.
$\vec E_j(\vec r)e^{i\vec k_j\cdot\vec r-i\omega_jt}$
denotes the running wave amplitude of the laser field.

We define $\Delta=\omega_1-\omega_{eg}$ and
$\delta_j=\omega_j-\omega_1$. Without loss of generality,
assuming $\vec{d}_{eg}=\vec{d}_{eg'}
=\vec{d}$, $\Omega_j=2\vec{d}\cdot\vec{E}_j/\hbar$ is
the Rabi frequency for the corresponding dipole connected
transition. In the limit of a resonant two photon process with
large detuning from the intermediate atomic excited state
$\left|e\right\rangle$, we adiabatically eliminate
$\left|e\right\rangle$ to obtain an effective Hamiltonian
\begin{eqnarray}
{\cal H}_{\rm eff}=
\sum_{\mu=g,g'}\left[\frac{P^2}{2M}
+V_{\mu}(\vec{r})\right]\left|\mu\right\rangle\!\left\langle\mu\right|+V_{B/R},
\end{eqnarray}
with,
\begin{eqnarray}
V_B&=&\frac{\hbar}{4\Delta}\sum_j\Omega_je^{i\vec{k}_j\cdot\vec{r}-i\delta_jt}
\sum_j\Omega_j^*e^{-i\vec{k}_j\cdot\vec{r}+i\delta_jt}
\left|g\right\rangle\!\left\langle g\right|,\\
V_R&=&\frac{\hbar}{4\Delta}\Omega_1e^{i\vec{k}_1\cdot\vec{r}-i\omega_1t}
\Omega_2^*e^{-i\vec{k}_2\cdot\vec{r}+i\omega_2t}
\left|g\right\rangle\!\left\langle g'\right|+h.c., \hskip 24pt
\end{eqnarray}
where distinguish the two types Raman process according to
whether the final internal state
$\left|g'\right\rangle$ is the same as or different from the
initial state $\left|g\right\rangle$, respectively.
When $\left|g'\right\rangle=\left|g\right\rangle$, we identify
the general Raman process as Bragg diffraction \cite{kozuma99}.

In momentum space, the above two photon process involves the
simultaneous absorption of a photon
$\left(\vec{k}_1,\omega_1\right)$ by an atom in state
$\left|g;\vec{p}\right\rangle$ and stimulated transition to state
$\left|g';\vec{p}-\hbar\vec{k}_2+\hbar\vec{k}_1\right\rangle$ with
emission of a photon $\left(\vec{k}_2,\omega_2\right)$. For the
case of \textbf{Bragg diffraction} ($g' = g$) with two
counter-propagating waves
\begin{eqnarray}
\vec{k}_1&=&\vec{k},\nonumber\\
\vec{k}_2&\approx& -\vec{k},\nonumber\\
\vec{K}&=&\vec{k}_1-\vec{k}_2\approx 2\vec{k},
\end{eqnarray}
we obtain
\begin{eqnarray}\label{VBa}
V_B(\vec{r},t)&=&\frac{\hbar}{4\Delta}
\left[\left|\Omega_1\right|^2+\left|\Omega_2\right|^2\right.\nonumber\\
&&\left.+\Omega_1\Omega_2^*e^{i\delta_2t}e^{i\vec{K}\cdot\vec{r}}
+h.c.\right]
\left|g\right\rangle\!\left\langle g\right|.
\end{eqnarray}
The first two terms are AC Stark shifts that can be neglected. For
an atom initially in state $\left|g,\vec{p}_g\right\rangle$, the
last two terms of Eq.~(\ref{VBa}) couples to momentum states
$\vec{p}_g\pm 2\hbar\vec{k}$, respectively. By appropriately
choosing $\delta_2$ one can selectively enhance coupling to only
one momentum state or side mode. The resonance condition is
defined by energy conservation in the two photon process
\begin{eqnarray}
\hbar\omega_1+\frac{\vec{p}_g^2}{2M}
\equiv\hbar\omega_2+\frac{\left(\vec{p}_g+\hbar\vec{K}\right)^2}{2M},
\end{eqnarray}
which gives
\begin{eqnarray}\label{reson}
\hbar\delta_2=\frac{\vec{p}_g^2}{2M}
-\frac{\left(\vec{p}_g+\hbar\vec{K}\right)^2}{2M}\approx
-\frac{\hbar^2\vec{K}^2}{2M},
\end{eqnarray}
for an atom with $\vec{p}_g\sim 0$. Hence under conditions
resonant for Bragg diffraction, the effective coupling (\ref{VBa})
creates superposition states like,
\begin{eqnarray}\label{pB1}
|\psi\rangle=\alpha\left|g\right\rangle
\left|\vec{p}_g\right\rangle+\beta\left|g\right\rangle\left|\vec{p}_g\pm
2\hbar\vec{k}\right\rangle.
\end{eqnarray}

For Bose-condensed atoms, the above discussion still applies to
any of the identical single atoms. When a condensate of N atoms is
involved (initial motional state $\left|p_g \sim 0\right\rangle$),
one simply creates a condensate in state (\ref{pB1}), which is not
an entangled state because it is simply putting N atoms into the
same state (\ref{pB1}), i.e.
\begin{eqnarray}
\left(a_{\left|\psi\right\rangle}^\dag\right)^N
\left|{\rm vac}\right\rangle\sim\psi(\vec{r}_1)\psi(\vec{r}_2)\cdots\psi(\vec{r}_N)
=\left|\psi\right\rangle^{\otimes N}.
\end{eqnarray}
In practice this can only be done if atoms are noninteracting, or
that the effect of interactions is small during the time of atomic
Bragg diffraction. Bragg diffraction of condensate atoms was
demonstrated first by the NIST-Gaithersburg group \cite{kozuma99}.
Higher motional states $\vec{K} = \pm 2m\vec{k}$ ($m$ integer) was
also obtained through, higher order, $2m$-photon couplings.

For the \textbf{Raman process} with two co-propagating waves
$\vec{k}_1\approx \vec{k}_2$ involving nearly degenerate ground
states $\left|g\right\rangle$ and $\left|g'\right\rangle$, we have
\begin{eqnarray}\label{VRa}
V_R=\frac{\hbar}{4\Delta}\Omega_1\Omega_2^*e^{i\delta_2 t}\left|g\right\rangle\!\left\langle
g'\right|+h.c.,
\end{eqnarray}
where AC Stark shifts from second order processes involving the
same laser fields have been neglected. The Raman resonance
condition is then simply $\omega_1-\omega_2 = \omega_{g'g}$. Note
that co-propagating Raman results in an effective coupling
(\ref{VRa}) which is constant, independent of the position
$\vec{r}$. The two photon process then is essentially
insensitive to motional effects $(\vec k_2\sim\vec k_1)$.

\section{Two-atom, two-photon process through an intermediate molecular state}
We now consider a model involving two $\Lambda$-type atoms whose
initial and final states are described by the same non-interacting
atomic states. The intermediate excited state, however, is now
chosen to be a bound, molecular, excited state, similar to those
utilized in recent photo-association experiments \cite{wynar00}.
Neglecting configurations not directly involved in the two photon
process, the two atom Hamiltonian can be written as
\begin{widetext}
\begin{eqnarray}\label{H2}
{\cal H}&&=\sum_{\mu=g,g'}\left[\sum_{i=1,2}
\left(\frac{\vec{p}_i^2}{2M}+V_{\mu}(\vec{r}_i)\right)+2\hbar\omega_{\mu
g}\right]\left|\mu,\mu\right\rangle\!\left\langle\mu,\mu\right|
+\sum_{\mu=g,g'}U_{\mu\mu}(\left|\vec{r}_1-\vec{r}_2\right|)
\left|\mu,\mu\right\rangle\!\left\langle\mu,\mu\right|\nonumber\\
&&+\left[\sum_{i=1,2}\left(\frac{\vec{p}_i^2}{2M}
+V_{g'g}(\vec{r}_i)\right)+U_{g'g}(\left|\vec{r}_1-\vec{r}_2\right|)
+\hbar\omega_{g'g}\right]\left|g',g\right\rangle\!\left\langle
g',g\right|\nonumber\\
&&+\left[\sum_{i=1,2}\left(\frac{\vec{p}_i^2}{2M}+V_{e_b}(\vec{r}_i)\right)
+U_{e_b}(\left|\vec{r}_1-\vec{r}_2\right|)+\hbar\omega_{eg}\right]
\left|e_b\right\rangle\!\left\langle
e_b\right|+H_{{\rm int}},
\end{eqnarray}
\end{widetext}
where states $\left|\mu,\mu\right\rangle$ for $\mu = g,g'$ denote
symmetrized electronic states of two atoms, each in
$\left|\mu\right\rangle$. $\left|g',g\right\rangle$ denote the
symmetrized state with one in $\left|g\right\rangle$ and the other
in $\left|g'\right\rangle$. $V_{\mu = g,g',g'g,e_b}$ denote
external trapping potentials in state $\left|\mu\right\rangle$,
$U_{\mu\mu}$ and $U_{g'g}$ are atom-atom interactions which at
large inter-atomic distances are described by the usual van der
Waals type terms, and at short range modified by coulomb effects.
The intermediate molecular state family, denoted by
$\left|e_b\right\rangle$, can be accessed through the direct
dipole coupling term $H_{{\rm int}}$. $U_{e_b}$ is the
Born-Oppenheimer molecular (of atom 1 and 2) binding potential for
internal state manifold $\left|e_b\right\rangle$, which contain
bound levels to be used as intermediate states.

Introducing center of mass $\vec{R} = (\vec{r}_1+\vec{r}_2)/2$ and
relative coordinate $\vec{r} = \vec{r}_2 - \vec{r}_1$ for the two
atoms, we can express
\begin{eqnarray}
\vec{r}_1=\vec{R}-\frac{\vec{r}}{2},&&
\vec{r}_2=\vec{R}+\frac{\vec{r}}{2},
\end{eqnarray}
and
\begin{eqnarray}
\frac{\vec{p}_1^2}{2M}+\frac{\vec{p}_2^2}{2M}=\frac{\vec{P}^2}{2(2M)}+\frac{\vec{p}^2}{2(M/2)}.
\end{eqnarray}

When $\left|e_b\right\rangle$ is asymptotically connected with
$\left|e,\mu\right\rangle=\left(\left|e\right\rangle\left|\mu\right\rangle
+ \left|\mu\right\rangle\left|e\right\rangle\right)/\sqrt{2}$ (for
$\mu = g,g'$), we can express the dipole coupling as
\begin{eqnarray}
\label{Hint2}
H_{{\rm int}}=
-\sum_{i=1,2}\vec{d}_i\cdot\vec{E}_1e^{i\vec{k}_1\cdot\vec{r}_i-i\omega_1t}
\left(\left|g\right\rangle\!\left\langle e\right|\right)_i+h.c.\nonumber\\
-\sum_{i=1,2}\vec{d}_i\cdot\vec{E}_2e^{i\vec{k}_2\cdot\vec{r}_i-i\omega_2t}
\left(\left|g'\right\rangle\!\left\langle e\right|\right)_i+h.c.\nonumber\\
=-\frac{\hbar\Omega_1}{\sqrt{2}}\cos(\vec{k}_1\cdot\vec{r}/2)e^{i\vec{k}_1\cdot\vec{R}-i\omega_1t}
\left|g,g\right\rangle\!\left\langle e_b\right|+h.c.\nonumber\\
-\frac{\hbar\Omega_2}{\sqrt{2}}\cos(\vec{k}_2\cdot\vec{r}/2)e^{i\vec{k}_2\cdot\vec{R}-i\omega_2t}
\left|g',g'\right\rangle\!\left\langle e_b\right|+h.c.,
\end{eqnarray}
where we have assumed $\vec{d}_i = \vec{d}_i' = \vec{d}$ for the
$\left|\mu\right\rangle \rightarrow \left|e\right\rangle$
transition of atom $i$. We note that the electronic dipole
coupling $\Omega_j$ will vary considerably with $\vec{r}$ for
small values of $r$, due to the relatively short range of the
molecular interactions.

To express the intermediate state in terms of the particular
molecular resonance or bound state $\left|e_b\right\rangle$, we
now take a look at the eigen-structure within the
$\left|e_b\right\rangle$ family. We will essentially base our
discussion on some kind of molecular state asymptotically
connected to the limit of one ground $\left|g\right\rangle$ and
one excited atom $\left|e\right\rangle$, just as the $O_g^-$
state, extensively discussed by D. Heinzen and others in their
photo association work \cite{boesten99}. The recent experiment by
Heinzen and coworkers on the production of ground state molecules
from an atomic condensate by a two-photon Raman process
\cite{wynar00} provides additional motivation to explore our ideas
experimentally. The photo-association process
\cite{boesten99,bohn96,kostrun00} used by Heinzen \textit{et al.}
relies on the transition strength of going from a (two atom)
'free' to a (molecular) `bound' state via an intermediate, excited
(molecular) `bound' state. What we desire, on the other hand, is a
transition from a (two atom) `free' to 'free' (relative motion)
state via an intermediate, excited (molecular) 'bound' state. More
detailed discussions about feasibility of such transitions can be
given provided all molecular potential curves are available.

The question of whether this is possible or not does not seem to
depend on the sample density (in the weakly interacting limit),
but will depend, as we shall show below, to a large degree on the
trap strength and the excited bound state structure. Assuming all
trapping potentials $V_g$, $V_{g'}$, and $V_{e_b}$ to be harmonic,
the separation of center-of-mass and relative coordinates for the
trapping potentials is
\begin{eqnarray}
&&\frac{1}{2}M\sum_{j = 1,2,3}\nu_j^2x_{1j}^2+\frac{1}{2}M\sum_{j = 1,2,3}\nu_j^2x_{2j}^2\nonumber\\
&=&\frac{1}{2}\left(2M\right)\sum_{j=1,2,3}\nu_j^2X_{j}^2+\frac{1}{2}\left(M/2\right)\sum_{j = 1,2,3}\nu_j^2x_{j}^2\nonumber\\
&=&V_{tR}+V_{tr},
\end{eqnarray}
where $x_{ij} = \hat{e}_j\cdot\vec{r}_i$, $X_{ij} =
\hat{e}_j\cdot\vec{R}$, and $x_j = \hat{e}_j\cdot\vec{r}$ with
$\hat{e}_j$ the unit vector in the $j$ direction.

When short range interactions $U_{\mu\mu}(\vec{r})$ are
approximated by their optical contact forms, Wilkens and coworkers
provide analytic solutions for two interacting atoms inside a
harmonic trap \cite{busch98}. In principle, from
\begin{eqnarray}
\left[\frac{p^2}{2(M/2)}+V_{tr}^\mu(\vec{r})
+U_{\mu\mu}(\vec{r})\right]\phi_n^\mu(\vec{r})=E_n^\mu\phi_n^\mu(\vec{r}),
\end{eqnarray}
we can find all bound states $\phi_n^\mu(\vec{r})$ due to the
external trapping potential $U_{\mu\mu}(\vec{r})$. We note that if
the range of the inter-atomic interaction $\ll r \ll$ size of the
ground state of the harmonic trap, then these bound states should
resemble the standard low energy scattering solutions.

Similarly, the bound state of the intermediate, excited molecular
level satisfies
\begin{eqnarray}
\left[\frac{p^2}{2(M/2)}+V_{tr}^{e_b}(\vec{r})
+U_{e_b}(\vec{r})\right]\phi_m^{e_b}(\vec{r})=E_m^{e_b}\phi_m^{e_b}(\vec{r}),
\end{eqnarray}
where the energy $E_m^{e_b}$ is measured from the asymptotic
energy $\hbar\omega_{eg}$. The index $m$ represents the rotational
and vibrational quantum numbers. For all bound states with
$E_m^{e_b} < 0$, the effect of external trapping potential
$V_{tr}^{e_b}$ is much less than for the free atoms, since the
corresponding bound state wave function is localized inside the
molecular potential well $U_{e_b}$.

Similarly, complete bound states $\left|l\right\rangle$ can also
be introduced for $\left|g',g\right\rangle$ if needed. They
satisfy
\begin{eqnarray}
\left[\frac{p^2}{2(M/2)}+V_{tr}^{g'g}(\vec{r})
+U_{g'g}(\vec{r})\right]\psi_l(\vec{r})=E_l^{g'g}\psi_l(\vec{r}).
\end{eqnarray}

In the $(\vec{R},\vec{r})$ basis, keeping only one intermediate
bound state $m_b$, we can then write our previous Hamiltonian,
Eq.~(\ref{H2}), as
\begin{widetext}
\begin{eqnarray}
{\cal H}&=&\sum_{n_\mu}\sum_{\mu=g,g'}
\left[\frac{\vec{P}^2}{2(2M)}+V_{tR}^\mu(\vec{R})+2\hbar\omega_{\mu g}
+E_n^\mu\right]
\left|\mu,\mu;n_\mu\right\rangle\!\left\langle\mu,\mu;n_\mu\right|\nonumber\\
& &
+\sum_{l}\left[\frac{\vec{P}^2}{2(2M)}+V_{tR}^{g'g}(\vec{R})
+\hbar\omega_{g'g}+E_l^{g'g}\right]\left|g',g;l\right\rangle\!\left\langle
g',g;l\right|\nonumber\\
&&+\left[\frac{\vec{P}^2}{2(2M)}+V_{tR}^\mu(\vec{R})
+\hbar\omega_{eg}+E_{m_b}^{e_b}\right]\left|e_b;m_b\right\rangle\!\left\langle
e_b;m_b\right|+H_{{\rm int}},
\end{eqnarray}
and the interaction term Eq.~(\ref{Hint2}) now becomes
\begin{eqnarray}
H_{{\rm int}} &=&
-\frac{\hbar\Omega_1}{\sqrt{2}}\sum_{n_g}\eta_{n_gm_b}e^{i\vec{k}_1\cdot\vec{R}
-i\omega_1t}\left|g,g;n_g\right\rangle\!\left\langle e_b;m_b\right|+h.c.\nonumber\\
&&-\frac{\hbar\Omega_2}{\sqrt{2}}\sum_{n_{g'}}\eta_{n_{g'} m_b}
e^{i\vec{k}_2\cdot\vec{R}/2-i\omega_2t}
\left|g',g';n_{g'}\right\rangle\!\left\langle e_b;m_b\right|+h.c.
,
\end{eqnarray}
\end{widetext}
where
\begin{eqnarray}
\eta_{n_\mu m_b}&=&{1\over |\vec d|}\left\langle
n_\mu\right|d(\vec{r})\cos(\vec{k}\cdot\vec{r}/2)\left|m_b\right\rangle\nonumber\\
&\approx &\frac{1}{d}\int
d\vec{r}\phi_n^{\mu*}(\vec{r})d(\vec{r})\cos(\vec{k}\cdot\vec{r}/2)\phi_{m_b}^{e_b}(\vec{r}),
\end{eqnarray}
which is a measure of the free-bound transition amplitude. we note
that in contrast to other expressions for the free-bound
transition amplitude \cite{boesten99,bohn96,kostrun00}, the
motional wave functions $\left|n_\mu\right\rangle$ and
$\left|m_b\right\rangle$ are now normalized in the trap rather
than energy or momentum normalized. Although the modification of
the inner part of the trap potential for $r \leq 100$ ($a_0$) due
to the inter-atomic potentials is to be included, these values can
be determined for selected molecular states from the results of
photo-association experiments. They can also be computed directly
if accurate potentials are available. Detailed discussions are
given by Verhaar \cite{boesten99}, Julienne \cite{bohn96} and
Javaninen \cite{kostrun00}.

In general, the coupling coefficient will vary depending on
intermediate state. It is therefore important to pick the largest
one of them. We then perform an adiabatic elimination of the
excited state by appropriate choice of laser detunings, and obtain
the effective Hamiltonian,
\begin{widetext}
\begin{eqnarray}
{\cal H}&=&\sum_{n_\mu}\sum_{\mu=g,g'}\left[\frac{\vec{P}^2}{2(2M)}
+V_{tR}^\mu(\vec{R})+2\hbar\omega_{\mu g}+E_n^\mu\right]
\left|\mu,\mu;n_\mu\right\rangle\!\left\langle\mu,\mu;n_\mu\right|\nonumber\\
&
&+\sum_{l}\left[\frac{\vec{P}^2}{2(2M)}+V_{tR}^{g'g}(\vec{R})
+\hbar\omega_{g'g}+E_l^{g'g}\right]\left|g',g;l\right\rangle\!\left\langle
g',g;l\right|+V_{B/R},
\end{eqnarray}
with
\begin{eqnarray}
V_B&&=\frac{\hbar}{4\Delta}2\sum_{n,l}
\left[\sum_j\Omega_j\eta_{nm_b}e^{i\vec{k}_j\cdot\vec{R}-i\omega_jt}\right]
\left[\sum_j\Omega_j\eta_{lm_b}e^{i\vec{k}_j\cdot\vec{R}-i\omega_jt}\right]^\dag
\left|g,g;n\right\rangle\!\left\langle
g,g;l\right|,\nonumber\\
V_R&&=\frac{\hbar}{4\Delta}2\sum_{n,n'}
\left[\Omega_1\eta_{nm_b}e^{i\vec{k}_j\cdot\vec{R}-i\omega_1t}\right]
\left[\Omega_2\eta_{n'm_b}e^{i\vec{k}_2\cdot\vec{R}-i\omega_2t}\right]^\dag
\left|g',g';n'\right\rangle\!\left\langle
g,g;n\right|+h.c.,
\end{eqnarray}
\end{widetext}
where we have defined $\Delta = \omega_1 - \omega_2 -
E_{n_b}^{e_b}/\hbar$. Note the $(\sqrt{2})^2$ enhancement due to
excitation to an asymptotic, symmetric state.

More generally, one can include all motional states and still be
able to find the effective coupling between the selected
electronic states, to obtain similar results \cite{you01}.

\begin{figure}
\includegraphics[width=3.in]{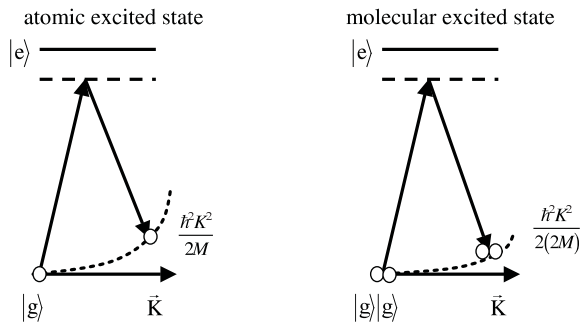}%
\caption{The illustration of
energy level diagrams for Bragg diffractions through
atomic (left part) and molecular excited states (right part).}
\label{fig0}
\end{figure}

Our analysis now follows along the same lines as in the previous
case of a single atom. For example, if we consider only the lowest
motional state $n_\mu = 0$ (a situation well approximated by Bose
condensed atoms), then our Hamiltonian becomes
\begin{widetext}
\begin{eqnarray}
{\cal H}=\sum_{\mu=g,g'}\left[\frac{\vec{P}^2}{2(2M)}
+V_{tR}^\mu(\vec{R})+2\hbar\omega_{\mu
g}+E_m^\mu\right]\left|\mu,\mu;0\right\rangle\!\left\langle\mu,\mu;0\right|
+V_{B/R},
\end{eqnarray}
with
\begin{eqnarray}
V_B&=&\frac{\hbar}{2\Delta}\eta_{0m_b}\eta_{0'm_b}^*
\left[\sum_j\Omega_je^{i\vec{k}_j\cdot\vec{R}-i\omega_jt}\right]
\left[\sum_j\Omega_je^{i\vec{k}_j\cdot\vec{R}-i\omega_jt}\right]^\dag
\left|g,g;0\right\rangle\!\left\langle
g,g;0'\right|,\nonumber\\
V_R&=&\frac{\hbar}{2\Delta}\eta_{0m_b}\eta_{0'm_b}^*\Omega_1\Omega_2^*
\left[e^{i\vec{k}_j\cdot\vec{R}-i\omega_1t}\right]
\left[e^{i\vec{k}_2\cdot\vec{R}-i\omega_2t}\right]^\dag
\left|g',g';0\right\rangle\!\left\langle
g,g;0'\right|+h.c.
\end{eqnarray}

For \textbf{Bragg diffraction} by two counter-propagating waves,
the above result reduces to
\begin{eqnarray}
V_B(\vec{R},t)=\frac{\hbar}{2\Delta}\left|\eta_{0m_b}\right|^2
\left[\left|\Omega_1\right|^2+\left|\Omega_2\right|^2
+\Omega_1\Omega_2^*e^{i\delta_2t}e^{i\vec{K}\cdot\vec{R}}
+\Omega_1^*\Omega_2e^{-i\delta_2t}e^{-i\vec{K}\cdot\vec{R}}\right]
\left|g,g;0\right\rangle\!\left\langle
g,g;0\right|.
\end{eqnarray}
\end{widetext}
Similar to the single atom case, Bragg diffraction involves the
simultaneous absorption and stimulated emission of photons, i.e.
the elementary Bragg process involves a total of two photons.
In contrast, however, to the single atom case, a pair of atoms are
now diffracted as illustrated in Fig. \ref{fig0}.
Hence for Bose condensed atoms with $p_i \sim 0$,
Bragg diffraction produces atoms in the state
\begin{eqnarray}
\left|\psi\right\rangle_M
&=&\alpha\left|g,g;0\right\rangle\left|p_1\approx 0,p_2\approx 0\right\rangle
+\beta\left|g,g;0\right\rangle e^{\pm
i\vec{K}\cdot\vec{R}}\nonumber\\
&=&\alpha\left|g,g\right\rangle\left|p_1\approx 0,p_2\approx 0\right\rangle\nonumber\\
&&+\beta\left|g,g\right\rangle
\left|\vec{p}_1=\hbar\vec{K}/2,\vec{p}_2=\hbar\vec{K}/2\right\rangle.
\end{eqnarray}
Not surprisingly, the momentum shift is only half the value of
atomic Bragg diffraction, due to the fact that only two photons
are involved for the pair of atoms. The resonance condition occurs
at
\begin{eqnarray}
\hbar\omega_1+\frac{P_i^2}{2(2M)}+\frac{p_i^2}{2(M/2)}&=&\hbar\omega_2+\frac{P_f^2}{2(2M)}+\frac{p_f^2}{2(M/2)},\nonumber\\
p_i&\sim& p_f\sim 0,\nonumber\\
\vec{P}_f&=&\vec{P}_i\pm\hbar\vec{K},
\end{eqnarray}
which gives the resonance condition
\begin{eqnarray}
\hbar\delta_2=\frac{P_i^2}{2(2M)}-\frac{(\vec{P}_i\pm\hbar\vec{K})^2}{2(2M)}
\approx -\frac{\hbar^2\vec{K}^2}{2(2M)},
\end{eqnarray}
i.e. half the atomic Bragg resonance (\ref{reson}).

It is worth mentioning that for a condensate with many atoms, the
state obtained from diffraction via a two atom intermediate state
will be complicated. It is generally not of the form
$(a_{\left|\psi\right\rangle_M}^\dag)^{N/2}\left|{\rm vac}\right\rangle$.

For a nearly degenerate \textbf{Raman process} with two
co-propagating waves, we obtain for atoms initially at rest,
\begin{eqnarray}
V_R(\vec{R},t)&=&
{\hbar\over 2}\frac{\Omega_1\Omega_2^*}{\Delta}\eta_{0m_b}\eta_{0'm_b}^*\nonumber\\
&&e^{i\delta_2t}\left|g,g;0\right\rangle\!\left\langle g',g';0'\right|+h.c.
\end{eqnarray}
The resonance condition now becomes $\omega_1 - \omega_2 =
2\omega_{g'g}$, i.e. the Raman process changes each atom's state
($g \rightarrow g'$) and so each atom acquires the energy deficit
$\hbar\omega_{g'g}$. Note that there is no $\vec{R}$ dependence of
the coupling coefficient in this case.

If the Raman process is arranged to allow for simultaneous
scattering to two different final states $\left|g'\right\rangle$
and $\left|g''\right\rangle$, the coherent coupling becomes,
\begin{eqnarray}
V_R^{[2]}(\vec{R},t)&=&
\frac{\hbar\Omega_R}{2}e^{i\delta_2t}
\left|g,g;0\right\rangle\!\left\langle g',g';0'\right|+h.c.\nonumber\\
&&+\frac{\hbar\Omega_R'}{2}e^{i\delta_2't}
\left|g,g;0\right\rangle\!\left\langle g'',g'';0''\right|+h.c. \hskip 24pt
\end{eqnarray}
Such a coupling can be obtained among different Zeeman levels in
the atomic ground state.

Finally, we note that our discussions above can also be applied to
a pair of different species of atoms, e.g. using molecular
intermediate state from Li-Cs dimer could create entangled pairs
of Li and Cs. We also note that to be formally correct, the second
order perturbation process also needs to be considered for the
state $\left|g',g\right\rangle$, which should have some type of
effective couplings with itself and as well as with the states
$\left|g,g\right\rangle$ and $\left|g',g'\right\rangle$. we have
neglected those couplings, assuming that resonance conditions for
their survival in the interaction picture are not satisfied.

\section{Spin-spin interaction models and two mode approximations}
In this section, we will be primarily concerned with process which
produce entanglement of atomic spin states. We therefore consider
a spin 1/2 system and adopt the notation
$\left|\uparrow\right\rangle$ for spin up and
$\left|\downarrow\right\rangle$ for spin down, with the following
operators and identities:
\begin{eqnarray}
\sigma_z&=&\left|\uparrow\right\rangle\!\left\langle\uparrow\right|-\left|\downarrow\right\rangle\!\left\langle\downarrow\right|,\nonumber\\
\sigma_+&=&\frac{1}{2}(\sigma_x+i\sigma_y)=\left|\uparrow\right\rangle\!\left\langle\downarrow\right|,\nonumber\\
\sigma_-&=&\frac{1}{2}(\sigma_x-i\sigma_y)=\left|\downarrow\right\rangle\!\left\langle\uparrow\right|,\nonumber\\
\sigma_+^2&=&\sigma_-^2=0,\nonumber\\
\sigma_+\sigma_-+\sigma_-\sigma_+&=&1,\nonumber\\
\sigma_x^2&=&\sigma_y^2=\sigma_z^2=1.
\end{eqnarray}

The Bragg/Raman coupling as discussed here is of the form
\begin{eqnarray}
{\hbar\Omega_R\over 2}
\left[(|\!\uparrow\rangle\!\langle\downarrow\!|)_1\otimes
(|\!\uparrow\rangle\!\langle\downarrow\!|)_2
+(|\!\downarrow\rangle\!\langle\uparrow\!|)_1\otimes
(|\!\downarrow\rangle\!\langle\uparrow\!|)_2\right]\nonumber\\
={\hbar\Omega_R\over 2}(\sigma_+^{(1)}\otimes\sigma_+^{(2)}
+\sigma_-^{(1)}\otimes\sigma_-^{(2)}),
\end{eqnarray}
if we designate $\left|g\right\rangle$ and $\left|g'\right\rangle$
as $\left|\uparrow\right\rangle$ and
$\left|\downarrow\right\rangle$, respectively. The Rabi frequency
$\Omega_R$ is the two-photon Rabi frequency for transitions
between $\left|\uparrow\right\rangle$ and
$\left|\downarrow\right\rangle$ via the intermediate molecular
state. Hence it contains the single photon Rabi frequencies
$\Omega_1$ and $\Omega_2$, as well as the free-bound transition
amplitude, $\eta_{n_\mu m_b}$. This is a new type of coupling not
widely discussed before. Its ability, however, to generate
entanglement should be obvious as two atoms perform conditional
evolution at all times, similar to the coupling for photon down
conversion. Its prospects for creating massive GHZ states will now
be studied.

We first compare our coupling scheme with other relevant models.
In the original scheme of Sorensen and Molmer \cite{sorensen99},
the two atom coupling takes the following form,
\begin{eqnarray}
&&\frac{\hbar\Omega_R}{2}\sigma_x^{(1)}\otimes\sigma_x^{(2)}\nonumber\\
&&=\frac{\hbar\Omega_R}{2}
\left(\left|\uparrow\right\rangle\!\left\langle\downarrow\right|
+\left|\downarrow\right\rangle\!\left\langle\uparrow\right|\right)_1\otimes
\left(\left|\uparrow\right\rangle\left\langle\downarrow\right|
+\left|\downarrow\right\rangle\!\left\langle\uparrow\right|\right)_2,
\end{eqnarray}
a form different from ours. It is convenient to analyze such
models for the case of many atoms in terms of collective spin
operator $J_{\mu = x,y,z} = \frac{1}{2}\sum_{i}\sigma_\mu^{(i)}$.
One can then show that the Hamiltonian for the Sorensen and Molmer
scheme becomes
\begin{eqnarray}
\sum_{i<j}\frac{\hbar\Omega_R}{2}\sigma_x^{(i)}\otimes\sigma_x^{(j)}
\hbar\Omega_R\left(J_x^2-\frac{N}{4}\right),
\end{eqnarray}
while our model gives
\begin{eqnarray}
&&\sum_{i<j}\frac{\hbar\Omega_R}{2}\left(\sigma_+^{(i)}\otimes\sigma_+^{(j)}
+\sigma_-^{(i)}\otimes\sigma_-^{(j)}\right)\nonumber\\
&&=\frac{\hbar\Omega_R}{4}\left(J_+^2+J_-^2\right)\nonumber\\
&&=\frac{\hbar\Omega_R}{2}\left(J_x^2-J_y^2\right).
\end{eqnarray}

Recently, a many body, two mode coupling was proposed by Zoller
and coworker \cite{sorensen01}. They considered an interacting,
two component (i.e. $|\downarrow\rangle$ and $|\uparrow\rangle$)
condensate whose many body Hamiltonian becomes
\begin{eqnarray}
\label{Hzollerint}
H&=&\sum_{j=\downarrow,\uparrow}\epsilon_{0,j}\int
d\vec{r}\left|\phi_0(\vec{r})\right|^2a_j^\dag a_j\nonumber\\
&&+\frac{1}{2}\sum_{j=\downarrow,\uparrow}U_{jj}\int
d\vec{r}\left|\phi_0(\vec{r})\right|^4a_j^\dag a_j^\dag a_ja_j\nonumber\\
&&+U_{\downarrow\uparrow}\int
d\vec{r}\left|\phi_0(\vec{r})\right|^4a_\downarrow^\dag
a_\uparrow^\dag a_\uparrow a_\downarrow.
\end{eqnarray}
under the two mode approximation, where each component
has the same spatial mode,
\begin{eqnarray}
\Psi_j(\vec{r})=\phi_0(\vec{r})a_j.
\end{eqnarray}

Introducing the collective spin operators
\begin{eqnarray}
J_x&=&\frac{1}{2}\left(a_0^\dag a_1+a_1^\dag a_0\right),\nonumber\\
J_y&=&\frac{i}{2}\left(a_0^\dag a_1-a_1^\dag a_0\right),\nonumber\\
J_z&=&\frac{1}{2}\left(a_1^\dag a_1-a_0^\dag a_0\right),
\end{eqnarray}
and taking the index $0$ for $|\!\downarrow\rangle$ and $1$ for $|\!\uparrow\rangle$
\cite{spekkens99,milburn97,elgaroy99,yurke8604}, one can show that
$J_{x,y,z} = \frac{1}{2}\sum_{i}\sigma_{x,y,z}^{(i)}$ as was used
earlier. We also note that $N=a_1^\dag a_1+a_0^\dag a_0$ and the
Casimir relation
\begin{eqnarray}
J_x^2+J_y^2+J_z^2=\frac{N}{2}(\frac{N}{2}+1).
\end{eqnarray}
The interaction term Eq. (\ref{Hzollerint}) then becomes
\begin{eqnarray}
\frac{1}{2}J_z^2\left(U_{00}+U_{11}-2U_{01}\right)\int
d\vec{r}\left|\phi_0(\vec{r})\right|^4.
\end{eqnarray}

In the discussions of the numerical solutions to follow, we will
use the second quantized notation, and consider a pure initial
state with a fixed, total number of atoms $N$. Specifically, we
compare the three types of coupling,
\begin{eqnarray}\label{M1}
V_M=\hbar\Omega_RJ_x^2,
\end{eqnarray}
considered by Sorensen and Molmer \cite{sorensen99}, the spin
squeezing model
\begin{eqnarray}\label{S1}
V_S=\hbar\Omega_RJ_z^2,
\end{eqnarray}
recently considered by Zoller and coworkers \cite{sorensen01}, and
our proposed coupling
\begin{eqnarray}\label{br}
V_{B/R}=\frac{\hbar\Omega_R}{2}\left(J_x^2-J_y^2\right).
\end{eqnarray}

Given an initial state $\left|\psi(0)\right\rangle= a_1^{\dag
N}\left|{\rm vac}\right\rangle/\sqrt{N!}$, i.e. with all atoms
condensed in state $|\uparrow\rangle$, the time evolution operator
for $V_M$ is analytically known and takes a simple form at
$\Omega_Rt = \pi/2$ for $N$ an even integer \cite{sorensen99},
\begin{eqnarray}
U_M=e^{-iJ_x^2(\Omega_Rt)}=e^{i\pi
N/2+i\pi/4}\frac{1}{N!}a_0^{\dag N}a_1^N+e^{-i\pi/4}. \hskip 12pt
\end{eqnarray}
It creates the maximally entangled N-GHZ state
\begin{eqnarray}\label{ghzN}
\left|{\rm GHZ}\right\rangle_N&\propto &\frac{1}{\sqrt{2}}
\left[\left|0\right\rangle_0\left|N\right\rangle_1
+i^{N+1}\left|N\right\rangle_0\left|0\right\rangle_1\right].
\end{eqnarray}
In the context of Bose-Einstein condensation, state
$\left|{\rm GHZ}\right\rangle_N$
is an example
of the interesting fragmented condensate \cite{nozieres82}. The
Sorensen and Molmer model (Eq.~(\ref{M1})) produces perfect
GHZ-type states at selected times, while our coupling (Eq.~(\ref{br}))
doesn't, in general, produce exact GHZ-type state.
From numerical simulations,
however, we find that our model, on average, produces more than
50\% overlap with GHZ-type state (Eq.~(\ref{ghzN})) at selected
times.
Because of the geometric equivalence between $J_x^2$ and $J_z^2$,
the spin squeezing model (Eq.~(\ref{S1})) also produces
a perfect GHZ at selected times, except now the entanglement
is between single particle states
$|\pm\rangle=(|\uparrow\rangle\pm |\downarrow\rangle)/\sqrt{2}$ \cite{You}
and the initial state has to become $|\pm\rangle^{\otimes N}$.

For our model, we were not able to find a general analytic form
of the time evolution operator, even at specific values of $t$.
To numerically calculate the time evolution, we expand the wave
function as
\begin{eqnarray}
\left|\psi(t)\right\rangle=\sum_{m=0}^Nc_m(t)\left|N-m\right\rangle_0\left|m\right\rangle_1,
\end{eqnarray}
where
\begin{eqnarray}
\left|m\right\rangle_i=\frac{a_i^{\dag m}}{\sqrt{m!}}\left|{\rm vac}\right\rangle,
\end{eqnarray}
with the initial conditions given by the $c_m(0)$s.

\begin{figure}
\includegraphics[width=3.25in]{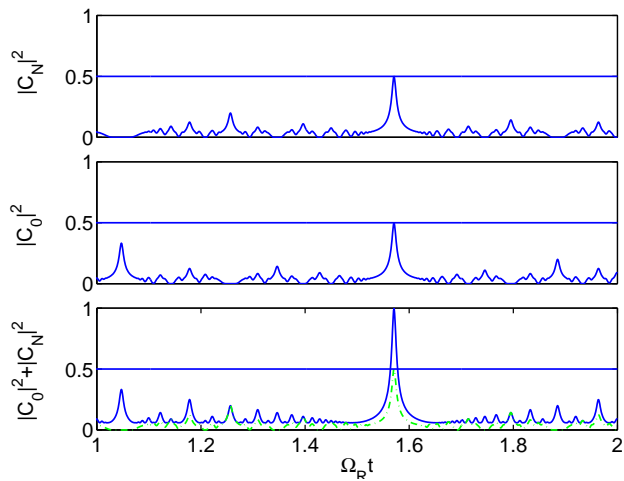}%
\caption{\label{fig1}The time dependent probabilities for being in
states $\left|0\right\rangle_0\left|N\right\rangle_1$ ($|c_N|^2$)
and $\left|N\right\rangle_0\left|0\right\rangle_1$ ($|c_0|^2$)
from the $V_M$ model.  The initial conditions are
$c_m(0)=\delta_{mN}$ ($N=500$). For clarity, $|c_N|^2$ (dashed line) is
superimposed into the plot of $|c_0|^2+|c_N|^2$. Note the perfect
$\left|{\rm GHZ}\right\rangle_N$ state at $\Omega_Rt=\pi/2$.}
\end{figure}

\begin{figure}
\includegraphics[width=3.25in]{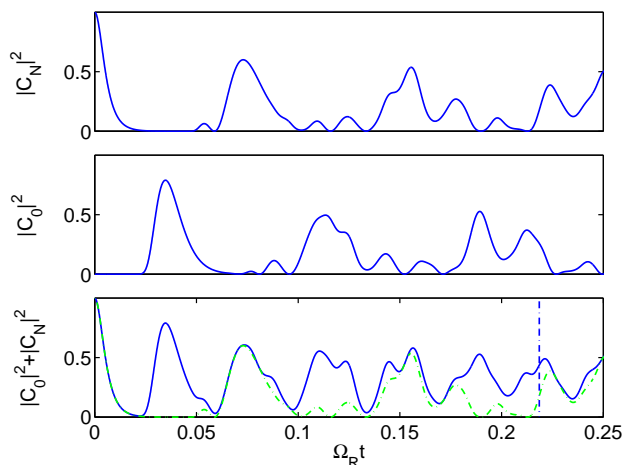}%
\caption{\label{fig2}The same as in Fig. \ref{fig1}
but for our model $V_{B/R}$.
Note that at several places the projection onto the
$\left|{\rm GHZ}\right\rangle_N$ state is already significant. At
$\Omega_Rt\sim 0.22$ (vertical line) the system exists in
$\left|{\rm GHZ}\right\rangle_N$ with about 50\% probability.}
\end{figure}

In Figs.~\ref{fig1} and \ref{fig2} we show, for comparison, the
generation of the maximum entangled GHZ state Eq.~(\ref{ghzN}) as
determined by model $V_M$ (Eq.~(\ref{M1})) and
our ``molecular diffraction" model (Eq.~(\ref{br})). We see that
while the Sorensen and Molmer model creates perfect massive GHZ
states at selected times, our model can also creates significant
overlap with the massive GHZ at selected times.

We can also compare the achievable spin squeezing between our
model and that of $V_S$ (Eq.~(\ref{S1})), using the
squeezing parameter
\begin{eqnarray}
\xi^2=\frac{N(\Delta J_{\vec{n}_1})^2}{\left\langle
J_{\vec{n}_2}\right\rangle^2+\left\langle
J_{\vec{n}_3}\right\rangle^2},
\end{eqnarray}
where $\vec{n}_i, i = 1,2,3$ are mutually orthogonal unit vectors.
Other discussions of spin squeezed states can be found in
\cite{kitagawa93,sorensen99b,hald99}.

For the initial state
\begin{eqnarray}\label{S10}
\left|\psi(0)\right\rangle=|+\rangle^{\otimes N}
=\frac{1}{2^{N/2}}\frac{1}{\sqrt{N!}}\left(a_0^\dag+a_1^\dag\right)^N\left|{\rm vac}\right\rangle,
\end{eqnarray}
it has been shown \cite{sorensen01} that
$\xi^2(t>0)<1$ for some set of $\vec{n}_i$'s. This model
of $V_S$ (Eq.~(\ref{S1})) is in fact the one-axis twisting model considered
by Kitagawa and Ueda earlier \cite{kitagawa93}. In this case the
problem can be solved analytically, with the result
\begin{eqnarray}
(\Delta J_{\vec{n}_1})^2_{{\rm min}}\sim N^{1/3}.
\end{eqnarray}
On the other hand, our model resembles the two-axis twisting model
of Kitagawa and Ueda \cite{kitagawa93}, and has to be solved
numerically. In the limit of large $N$ and with the condensate
initially in $|\uparrow\rangle^{\otimes N}$,
one can show that for an initial $J_z=N/2$ ($\vec J$-spin pointing along the
positive z-direction),
\begin{eqnarray}
(\Delta J_{\vec{n}_1})^2_{{\rm min}}\sim \frac{1}{2}.
\end{eqnarray}
The optimal squeezing in this case occurs with $J_x+J_y$. This
result can be easily verified by taking a semiclassical
approximation in the dynamical equation for $J_x$ and $J_y$. We
find that the time scale of reaching maximum squeezing is $\sim
1/(N\Omega_R)$ (see also \cite{wineland94}). For condensates
containing $10^6$ atoms, even with a very weak coupling $\Omega_R
\sim 1$ (Hz), the maximum squeezing is reached within a
micro-second.

\begin{figure}
\includegraphics[width=3.25in]{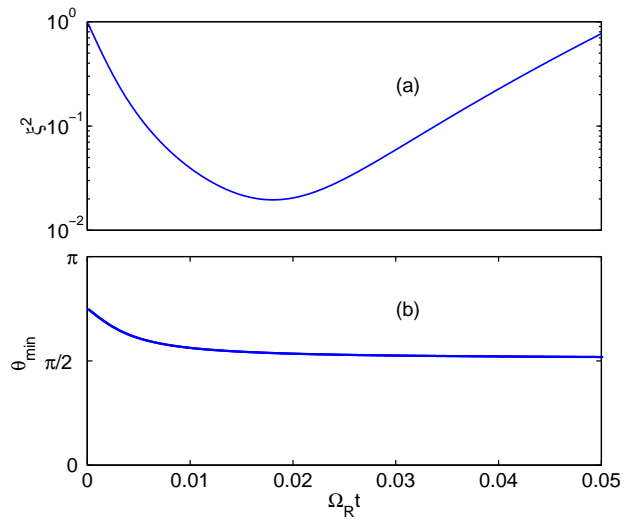}%
\caption{\label{fig3}
Numerical results from the model $V_S$.
(a) The time dependence of the minimal spin
squeezing parameter;
(b) The optimal angle when the minimal squeezing in (a) is achieved.}
\end{figure}

\begin{figure}
\includegraphics[width=3.25in]{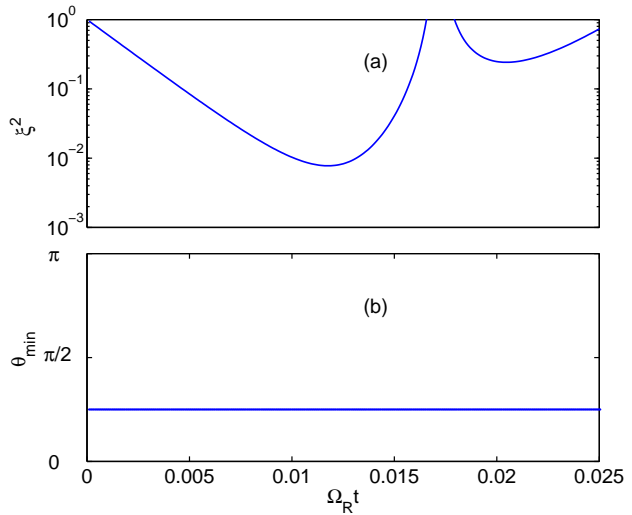}%
\caption{\label{fig4}
Same as in Fig. \ref{fig4} but from
our model $V_{B/R}$. Note that the minimal squeezing of
our model is smaller and the it occurs at a constant
spin direction.}
\end{figure}

In Figure \ref{fig3} the squeezing parameters and the
corresponding angle $\theta_{{\rm min}}$ are computed
numerically for the model $V_S$ (Eq.~(\ref{S1}))
with the initial state given by Eq.~(\ref{S10}). Fig.~\ref{fig4}
is a similar calculation for our model (Eq.~(\ref{br}))
starting with all atoms in state $|\uparrow\rangle$.
Our model achieves a better
squeezing at an earlier time. In addition, the phase angle
$\theta_{{\rm min}}$ is fixed at the value $\pi/4$. More
detailed discussions of the spin squeezing properties of our model
can be performed, including dissipation and finite system size
\cite{vardi01,law00,andre02}.

Finally we also note that our coupling scheme for simultaneous
transitions to two different final states gives
\begin{eqnarray}
V_{B/R}^{[2]}=\frac{\hbar\Omega_R}{4}\left[\left(a_0^\dag
a_0^\dag a_1a_1+a_0^\dag a_0^\dag a_{1'}a_{1'}\right)+h.c.\right].
\end{eqnarray}
In the undepleted pump limit when dissipation exists, or when the
coupling efficiency is small, this coupling can be approximated by
\begin{eqnarray}
V_{B/R}^{[2]}\approx {\cal E}\left(a_1^\dag
a_1^\dag+a_{1'}^\dag a_{1'}^\dag\right).
\end{eqnarray}
This is similar to the recent ideas of Refs.
\cite{pu0007,duan0011,vogels02} which creates entangled EPR pairs every
time an event occurs, just like in a parametric down conversion
process \cite{shih88,hong87}.

\section{Detection of the maximally entangled states}
While it seems that there are promising schemes to generating
massive entanglement of condensate atoms, the detection of
such entanglement represents a significant challenge by itself
as it is very difficult to perform individual atom resolved
measurements \cite{sackett00,wineland94} in a condensate.
Recently, it was suggested that performing a time reversed dynamic
evolution can detect the coherence of the created N-GHZ state
\cite{micheli02,vogels02}.

In this section, we detail the working mechanism
of the proposed scheme by performing explicit calculations.
We will use the $V_M$ model as an example.
To simplify the algebra, we will use $\tau=\Omega_R t$,
and denote the basis states according to
\begin{eqnarray}
\left|n_1\right\rangle&\equiv&\left|n_0,n_1\right\rangle
=\left|n_0\right\rangle_0\left|n_1\right\rangle_1.
\end{eqnarray}
Assume we start with $|\uparrow\rangle^{\otimes N}$,
then at $\tau=\pi/2$, a maximally entangled state
\begin{eqnarray}
|\psi(\tau={\pi\over 2})\rangle\propto {1\over
\sqrt{2}}(|N\rangle+i^{N+1}|0\rangle),
\label{ch}
\end{eqnarray}
Immediately reversing the time dynamics ($V_M\to -V_M$), it is
easy to show that after another $\tau=\pi/2$, we recover the
initial state $|\uparrow\rangle^{\otimes N}$, i.e. all atoms to be
detected are in state $|\uparrow\rangle$. If instead, the created
state at $\tau=\pi/2$ is a completely incoherent mixture of
$|0\rangle$ and $|N\rangle$, given by the (N-body) density matrix
\begin{eqnarray}
\rho(\tau={\pi\over 2})={1\over 2}|0\rangle\!\langle 0|+
{1\over 2}|N\rangle\!\langle N|,
\label{dm}
\end{eqnarray}
then the time reversed dynamics will lead to observations
of variable number of atoms in state $|\downarrow\rangle$ as well.

For the general case, let's assume the restricted form
\begin{eqnarray}
\rho(\tau={\pi\over 2})&=&p_0|0\rangle\!\langle 0|+
p_N|N\rangle\!\langle N|\nonumber\\
&&+\beta|0\rangle\!\langle N|
+\beta^*|N\rangle\!\langle 0|,
\end{eqnarray}
with $0\le p_{0/N}\le 1$, $p_0+p_N=1$,
and $|\beta|^2\le p_0p_N$.
Time reversed evolution of $V_M$ for $\tau$ from $\tau=\pi/2$
will lead to
\begin{eqnarray}
\rho(\tau+\pi/2)&=&U_M(-\tau)\rho(\pi/2)U_M^\dag (-\tau)\nonumber\\
&=&e^{iJ_x^2\tau}\rho(\pi/2)e^{-iJ_x^2\tau},
\end{eqnarray}
which can be explicitly evaluated with the use of
rotation operation
\begin{eqnarray}
e^{\pm iJ_x^2\tau}&=&e^{-i\frac{\pi}{2}J_y}e^{\pm iJ_z^2\tau}e^{i\frac{\pi}{2}J_y},
\end{eqnarray}
and the disentangling relation \cite{arecchi72}
\begin{eqnarray}
e^{-i\frac{\pi}{2}J_y}&=&e^{J_+}e^{J_z\ln{2}}e^{-J_-}=e^{-J_-}e^{-J_z\ln{2}}e^{J_+},\\
e^{i\frac{\pi}{2}J_y}&=&e^{-J_+}e^{J_z\ln{2}}e^{J_-}=e^{J_-}e^{-J_z\ln{2}}e^{-J_+}.
\end{eqnarray}

Denote
\begin{eqnarray}
A&=&\left\langle N\right|e^{J_+}e^{iJ_z^2\tau}e^{-J_+}\left|0\right\rangle,\\
B&=&\left\langle
N\right|e^{J_+}e^{iJ_z^2\tau}e^{J_-}\left|N\right\rangle,
\end{eqnarray}
then we find
\begin{eqnarray}\label{rhonn}
&&\rho_{NN}(\tau+\pi/2)=\langle N|\rho(\tau+\pi/2)|N\rangle\nonumber\\
&&={1\over 2^{2N}}
[\left|A\right|^2p_0+\left|B\right|^2p_N+(AB^*\beta+h.c.)]
\end{eqnarray}
with
\begin{eqnarray}
A&=&\sum_{m=0}^N\frac{N!}{m!(N-m)!}(-1)^me^{i(m-\frac{N}{2})^2\tau},\\
B&=&\sum_{m=0}^N\frac{N!}{m!(N-m)!}e^{i(m-\frac{N}{2})^2\tau}.
\end{eqnarray}
The following angular momentum algebra has been used,
\begin{eqnarray}
J_+\left|m\right\rangle&=&\sqrt{(m+1)(N-m)}\left|m+1\right\rangle,\nonumber\\
J_-\left|m\right\rangle&=&\sqrt{m(N-m+1)}\left|m-1\right\rangle,\nonumber\\
J_z\left|m\right\rangle&=&\left(m-\frac{N}{2}\right)\left|m\right\rangle,\nonumber\\
\left\langle m\right|J_+&=&(J_-\left|m\right\rangle)^\dag.
\end{eqnarray}
At $\tau=\frac{\pi}{2}$, we find
\begin{eqnarray}\label{rhonntau}
\rho_{NN}(\pi)=\frac{1}{2}+(-1)^{\frac{N}{2}}{\mathbf{Im}(\beta)}\le
1,
\end{eqnarray}
because $|{\mathbf{Im}(\beta)}|\leq\left|\beta\right|\leq
\sqrt{p_0p_N}\le 1/2$, for any state. Thus we conclude that the
probability of detecting atoms in state $|\downarrow\rangle$ is
nonzero for mixed states. In fact, for the completely incoherent
mixture Eq. (\ref{dm}) with $p_0=p_N=\frac{1}{2}$ and $\beta=0$,
$\rho_{00}(\pi)=\rho_{NN}(\pi)=1/2$.

\begin{figure}
\includegraphics[width=3.25in]{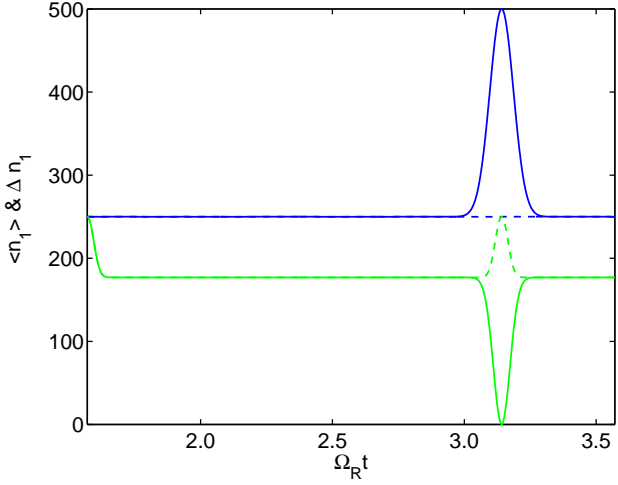}%
\caption{The time dependence of
average population
($n_1=a_1^\dag a_1$) and its variance ($\Delta n_1$)
for spin up $|\uparrow\rangle$ component.
The solid lines are for starting in the N-GHZ state Eq. (\ref{ch})
while the dashed lines are for starting in the incoherent
mixture Eq. (\ref{dm}). The upper two lines are
for $\langle n_1\rangle$ while the lower ones are for $\Delta n_1$.}
\label{fig5}
\end{figure}

It's interesting to compare the average populations
and fluctuations of spin up $|\uparrow\rangle$
number operator ($n_1=a_1^\dag a_1$)
after the time reversed evolution for
$\tau=\frac{\pi}{2}$ starting with the following
different initial conditions,
\begin{itemize}
\item A pure state, \\
$\left|\psi(\pi/2)\right\rangle=\frac{1}{\sqrt{2}}
(\left|N\right\rangle+i^{N+1}\left|0\right\rangle)$
    \begin{eqnarray}
        \left\langle n_1(\pi)\right\rangle&=&\mathbf{tr}(\rho
        n_1)=N,\nonumber\\
        \left\langle n_1^2(\pi)\right\rangle&=&\mathbf{tr}(\rho
        n_1^2)=N^2,\nonumber\\
        \Delta n_1(\pi)&=&\sqrt{\left\langle n_1^2\right\rangle-\left\langle
        n_1\right\rangle^2}=0;
    \end{eqnarray}
\item An incoherent mixture,\\
  $\rho(\pi/2)=\frac{1}{2}(\left|N\right\rangle\left\langle N\right|+\left|0\right\rangle\left\langle0\right|)$
    \begin{eqnarray}
        \left\langle n_1(\pi)\right\rangle&=&\mathbf{tr}(\rho
        n_1)=\frac{N}{2},\nonumber\\
        \left\langle n_1^2(\pi)\right\rangle&=&\mathbf{tr}(\rho
        n_1^2)=\frac{N^2}{2},\nonumber\\
        \Delta n_1(\pi)&=&\frac{N}{2}.
    \end{eqnarray}
\end{itemize}
Thus, these two cases become easily distinguishable. In Fig.
\ref{fig5}, we compare the time dependent results of
averaged population in state $|\uparrow\rangle$ and
its variance, from the time reversed evolution starting
with the coherent GHZ state Eq (\ref{ch})
and the incoherent mixture Eq. (\ref{dm}). It is
interesting to note that the average population of spin up
$|\uparrow\rangle$ number operator ($n_1=a_1^\dag a_1$) for the
completely incoherent mixture is in fact independent of $\tau$.
We find that the width of the Gaussian like features
near $\Omega_R t\sim \pi$ is inversely proportional to $/\sqrt{N}$,
but the plateau for the variance $\Delta n_1$
is at around $N/\sqrt{3}$.

For factorizable N-atom states, whether a pure state
$\left|\psi(\pi/2)\right\rangle
=\frac{1}{2^{N/2}}(\left|\uparrow\right\rangle
+\left|\downarrow\right\rangle)^{\otimes N}$
or a mixture
$\rho(\pi/2)=\frac{1}{2^N}
(\left|\downarrow\right\rangle\left\langle\downarrow\right|
+\left|\uparrow\right\rangle\left\langle\uparrow\right|)^{\otimes N}$,
the probability distribution after the time reversed
evolution gives identical results that are different from
the two multi-atom correlated states discussed earlier.
In fact, the final results simply
reflects the binomial distributions of $n_1$ in a
equal superposition separable state and
is given by
    \begin{eqnarray}
        \left\langle n_1(\pi)\right\rangle&=&\mathbf{tr}(\rho
        n_1)=\frac{N}{2},\nonumber\\
        \left\langle n_1^2(\pi)\right\rangle&=&\mathbf{tr}(\rho
        n_1^2)=\frac{N(N+1)}{4},\nonumber\\
        \Delta n_1(\pi)&=&\frac{\sqrt N}{2}.
    \end{eqnarray}
In this particular example with $V_M$ as the interaction,
the above results actually is independent of $\tau$.
Thus the measurement of total single particle state
population fluctuations after the
time reversed evolution provides a strong indication
for the coherence and correlation of the created entangled states.

\section{Limitations and advantages}
There are several limitations to our model that needs to be
discussed carefully. First, the molecular coupling needs to
dominate over the atomic coupling; i.e., we need to achieve a
molecular coupling
\begin{eqnarray}
\Omega_R^M&=&\frac{\Omega_1\Omega_2^*}{\Delta_M}\eta_{0m_b}\eta_{0'm_b}^*\nonumber\\
&\gg& \Omega_R^A = \frac{\Omega_1\Omega_2^*}{\Delta_A},
\end{eqnarray}
where $\Delta_{M/A}$ are detunings from the molecular and atomic
intermediate states, respectively. We can estimate analytically
the free-bound transition amplitude using the interacting ground
state wave function from \cite{busch98}, and excited state wave
function from the bound state solutions in a $-C_3/r^3$ potential
modified by an inner repulsive core \cite{gao99}. Most likely,
however, the deepest molecular bound state with significant
amplitude should be chosen in order to maximize the detuning from
the atomic transition. In addition, we should have $\Delta_M \gg
\gamma_M$ to minimize spontaneous emission which would lead to
decoherence and loss of atoms.

A clear advantage of our scheme is the signature for diffraction
via a molecular intermediate state. The diffracted atoms would
move at half the speed of atoms that have undergone atomic Bragg
diffraction. Even as the number of atoms is increased, our scheme
achieves the same level of squeezing and the same high value of
overlap with the massive GHZ state.

Our scheme is based on an engineered interaction that can be
turned on and off like the scheme of Sorensen and Molmer
(Eq.~(\ref{M1})). In particular, our scheme works for
non-interacting atoms which can decrease the noise due to
atom-atom interactions in a $U(1)$ symmetry breaking
condensate state. We have also provided explicit
calculations that demonstrate that the time reversed
dynamics can be used to verify the coherence of the
intended maximally entangled states.

Among all models discussed, there exists an even/odd number
problem: The form of the final state depends on whether the total number of
condensed atoms is even or odd number. If, however, the condensate
resembles a coherent superposition of number states (grand
canonical ensembles), we should average the results over the
number distribution $p(N)$ as well. This will be explored in a
future study.

This work benefited from our participation of the 2000 workshop
on quantum degenerate gases at the Lorentz Center, University of
Leiden. We thank H. Stoof for the hospitality. We thank Dr. M. S.
Chapman and S. L. Rolston for helpful discussions. This work is
supported by a grant from NSA, ARDA, and DARPA under ARO Contract
No. DAAD19-01-1-0667, and by a grant from the NSF PHY-0113831.

\end{document}